\documentclass[aps,prl,twocolumn,
 reprint, superscriptaddress,
 amsmath,amssymb,longbibliography
]{revtex4-2}
\usepackage[english]{babel}
 
\usepackage{dcolumn}
\usepackage{bm}

\usepackage{hyperref}
\hypersetup{
    colorlinks=true,
    linkcolor = [rgb]{0.70,0.13,0.13}, 
    citecolor = [rgb]{0.13,0.55,0.13},
    urlcolor = [rgb]{0.25, 0.41, 0.88} , 
    filecolor=cyan,      
    pdfcreator = {\LaTeX\ and \flqq hyperref\frqq},
}

\usepackage{cleveref}

\RequirePackage{graphicx,amsmath,amssymb,bm}
\usepackage{graphicx}
\usepackage{dcolumn}
\usepackage{bm}
\usepackage{appendix}
\RequirePackage{graphicx,amsmath,amssymb,bm}
\usepackage{bbm}
\usepackage{framed} 
\usepackage{amsthm}
\usepackage{caption}
\usepackage{subcaption}
\usepackage[T1]{fontenc}
\usepackage[utf8]{inputenc}
\usepackage[english]{babel}
\usepackage[normalem]{ulem}
\usepackage{comment}
\usepackage{latexsym}
\usepackage{amssymb}
\usepackage{amsmath}
\usepackage{amsfonts}
\usepackage[vcentermath]{youngtab}
\usepackage{upgreek}
\usepackage{bm,dsfont}
\usepackage{multirow}
\usepackage{braket}
\usepackage{enumitem}
\usepackage[percent]{overpic}
\usepackage[usenames, dvipsnames]{color}
\usepackage[svgnames]{xcolor}

\newcommand{\up}{\uparrow}
\newcommand{\dn}{\downarrow}

\graphicspath{ {./Figs/} }

 \newcommand{\e}{\text{e}}

\def\be{\begin{equation}}
\def\ee{\end{equation}} 
\def\bsh{\begin{shaded}}
\def\esh{\end{shaded}} 
\def\bpm{\begin{pmatrix}}
\def\epm{\end{pmatrix}}

\begin{document} 
\preprint{APS/123-QED}
\title{Universal Dynamic Scaling of 2D Quantum Ising Transition on the Fuzzy Sphere} 

\author{Meng Zeng} 
\email{mengzeng@pks.mpg.de}
\affiliation{Max Planck Institute for the Physics of Complex Systems, Nöthnitzer Str. 38, 01187 Dresden, Germany}
\author{Shuai Yin}
\affiliation{School of Physics, Sun Yat-Sen University, Guangzhou, China}
\affiliation{Guangdong Provincial Key Laboratory of Magnetoelectric Physics and Devices, Sun Yat-Sen University, Guangzhou, China}
\author{Roderich Moessner}
\affiliation{Max Planck Institute for the Physics of Complex Systems, Nöthnitzer Str. 38, 01187 Dresden, Germany}

\begin{abstract} 
We revisit the problem of \textit{real-time} quantum dynamics of the paradigmatic two dimensional transverse-field Ising model using the recently developed fuzzy sphere regularization scheme. By linearly ramping the transverse field from the paramagnetic phase to criticality, we study the finite-time scaling behavior of the squared order parameter $\langle m_z^2 \rangle$, the excitation energy density $Q$, and the two-point correlation function of $m_z$. We establish numerically that, at intermediate quench rate, $\langle m_z^2 \rangle$ follows the conventional Kibble-Zurek prediction set by the critical exponents of the $3$D Ising universality class, and the correlation function exhibits the expected exponential decay whose correlation length can be used to estimate the non-universal scaling coefficient in the freeze-out time/length. In contrast, the excitation energy density $Q$ does not reach the same scaling regime at available system sizes due to large effective finite-size gap from symmetry-enforced level sparsity in the energy spectrum.
At slow quench rates the universal quasi-adiabatic scaling for both $\langle m_z^2 \rangle$ and $Q$ is recovered. Since the 
fuzzy sphere construction can realize not only the Ising conformal field theory (CFT), but a broad family of $(2+1)d$ CFTs, our results establish a route to the real-time critical dynamics of 
strongly coupled CFTs that are otherwise computationally challenging to study. 
\end{abstract} 
\maketitle

\paragraph{\textcolor{violet}{Introduction}.}
The dynamics of many-body systems driven through a continuous phase transition is a fundamental questions in nonequilibrium physics. When the rate of the drive competes with the divergent relaxation time at criticality, the system inevitably falls out of equilibrium, and the resulting universal scalings of the topological defects are described by Kibble-Zurek (KZ) theory \cite{kibble1976topology,kibble1980some,zurek1985cosmological,zurek1996cosmological}. Then, based on the renormalization group theory, the full finite-time scaling (FTS) was developed to describe the dynamic scaling in the whole driven process for general physical quantities~\cite{Zhifangxu2005prb,Gong2010njp}.

KZ theory and the FTS form, also generalized to quantum phase transitions \cite{polkovnikov_colloquium_2011,jacek2010dynamics,dziarmaga_dynamics_2005,Zurek2005prl,Pol2005prl,review-heyl2018,heyl-scaling-universality-2015,damski_simplest_2005,federico-2025,sengupta_exact_2008,de_grandi_quench_2010,chandran_kibble-zurek_2012,Fischer2007prl,Yin2014open}, have been tested with remarkable precision in one spatial dimension, where exactly solvable models, density-matrix renormalization group, and a wealth of experimental platforms have provided quantitative confirmation across multiple universality classes  \cite{kolodrubetz_nonequilibrium_2012,heyl-dynamic-ising-transition-2013,heyl-broken-symmetry-2014,keesling_quantum_2019,heyl-observation-2017,dora_kibble-zurek_2019,lee-yang-2017,spacetime-2016,yin2026finite}. Physics beyond KZ has also been investigated \cite{zeng_universal_2023,beyond-KZ-roderich,deng2025anomalous}. In two and higher spatial dimensions, however, the situation is markedly less developed \cite{braun-emergence-coherence-2015,keesling_quantum_2019,du2023kibble}. Real-time simulation of strongly correlated quantum systems above one dimension remains a central computational obstacle: quantum Monte Carlo is carried out in imaginary time, requiring analytic continuation to extract real-time dynamics \cite{Sandvik2011prb,shu_equilibration_2025,weinberg2025defects,yin-dirac-2026,yin-dirac-natcomm-2025,shu2026universal}, while conventional tensor network methods are constrained by entanglement growth \cite{schmitt_quantum_2022,markus-2d-neuralnet}. As a result, controlled real-time dynamical tests of KZ theory beyond $1$d have been scarce \cite{schmitt_quantum_2022}. Very recently, however, Ref.~\cite{3d-ising-2026} managed to push the dynamics to $3$d using neural quantum states.

On another front, a recent development on the so-called fuzzy sphere regularization \cite{zhu_uncovering_2023} has dramatically advanced our ability to characterize strongly coupled CFTs in $(2+1)$ dimensions. In this construction, (spinful) particles are placed on the surface of a sphere with a magnetic monopole at the center, forming Landau levels. By carefully designing the Hamiltonian, various quantum phase transitions can be realized \cite{fuzzy-review,zhou_fuzzified_2025}. Crucially, the construction preserves exact $SO(3)$ rotational invariance already at the microscopic level, and the spherical geometry implements the state-operator correspondence, allowing direct access to the conformal spectrum at modest system sizes. Since its introduction, the method has provided high-precision results for the 3D Ising CFT \cite{zhu_uncovering_2023}, four-point functions and operator product expansion coefficients \cite{hu-ope-ising-2023,han-four-point-2025}, conformal defects and boundaries \cite{hu_solving_2024,zhou2025-surface-cft,dedushenko2024ising-ising-bcft,zohar-2024}, the entropic $F$-function \cite{f-function}, and has been extended to other universality classes \cite{han_conformal_2024,zhou_so5_2024,spN,dey2026conformal-O2,taige-ON,guo2025-ON,zhou2025free-maj,taylor2026conformal-ising-tricritical,zhou2025chern-simens,voinea2025-anyon,dey2026conformal-O3,fan2025simulating-lee-yang,potts-2025}. Yet despite this rapid progress, the fuzzy sphere has so far been confined to equilibrium, ground-state physics \cite{fuzzy-review}; its potential for nonequilibrium quantum critical dynamics is largely unexplored \cite{masud-2018}.

Our work aims to fill this gap. The guiding observation is simple: if the fuzzy sphere faithfully captures the equilibrium critical point, then KZ theory and its generalized FTS form — which are controlled entirely by the equilibrium critical theory and the quench protocol — should also be manifest in quench dynamics on the same geometry. One obvious advantage of the sphere geometry for this purpose is again the exact $SO(3)$ symmetry, which, with a symmetry-preserving quench protocol, restricts the dynamics to the same angular momentum sector as the initial state, sharply reducing the effective Hilbert space and enabling controlled tensor-network evolution to longer times than in conventional lattice geometries.

We test this expectation by ramping the fuzzy sphere Ising model linearly to its critical point at rate $v$, using a time-dependent variational principle (TDVP) algorithm \cite{tdvp} augmented by controlled bond expansion \cite{cbe}, complemented by exact diagonalization (ED) at smaller system sizes. We compute three observables at the critical point: the squared magnetization $\langle m_z^2\rangle$, the excitation energy density $Q$, and the two-point correlation function $C(\theta)$ as a function of the polar angle $\theta$. For $\langle m_z^2\rangle$ and $C(\theta)$ we find clean data collapse using the known 3D Ising critical exponents. Extrapolation to infinite system size confirms the predicted KZ power-law scaling of $\langle m_z^2\rangle$ with $v$. In contrast, for the excitation energy density $Q$, the scaling limit is not reached for the available systems due to much larger finite-size effects caused by the sparse spectrum from strong symmetry constraints. Furthermore, the slow-quench regime with long-time evolution accessible by ED recovers nicely the standard scaling behaviors in the adiabatic regime. Taken together, these results establish the fuzzy sphere as a quantitative laboratory for nonequilibrium quantum critical phenomena in (2+1) dimensions, including many other critical points realizable on the sphere \cite{fuzzy-review}. The methodology developed here in principle applies directly to all of them.

\paragraph{\textcolor{violet}{Fuzzy sphere setup and quench protocol}.}
\begin{figure}
    \centering
    \includegraphics[width=\linewidth]{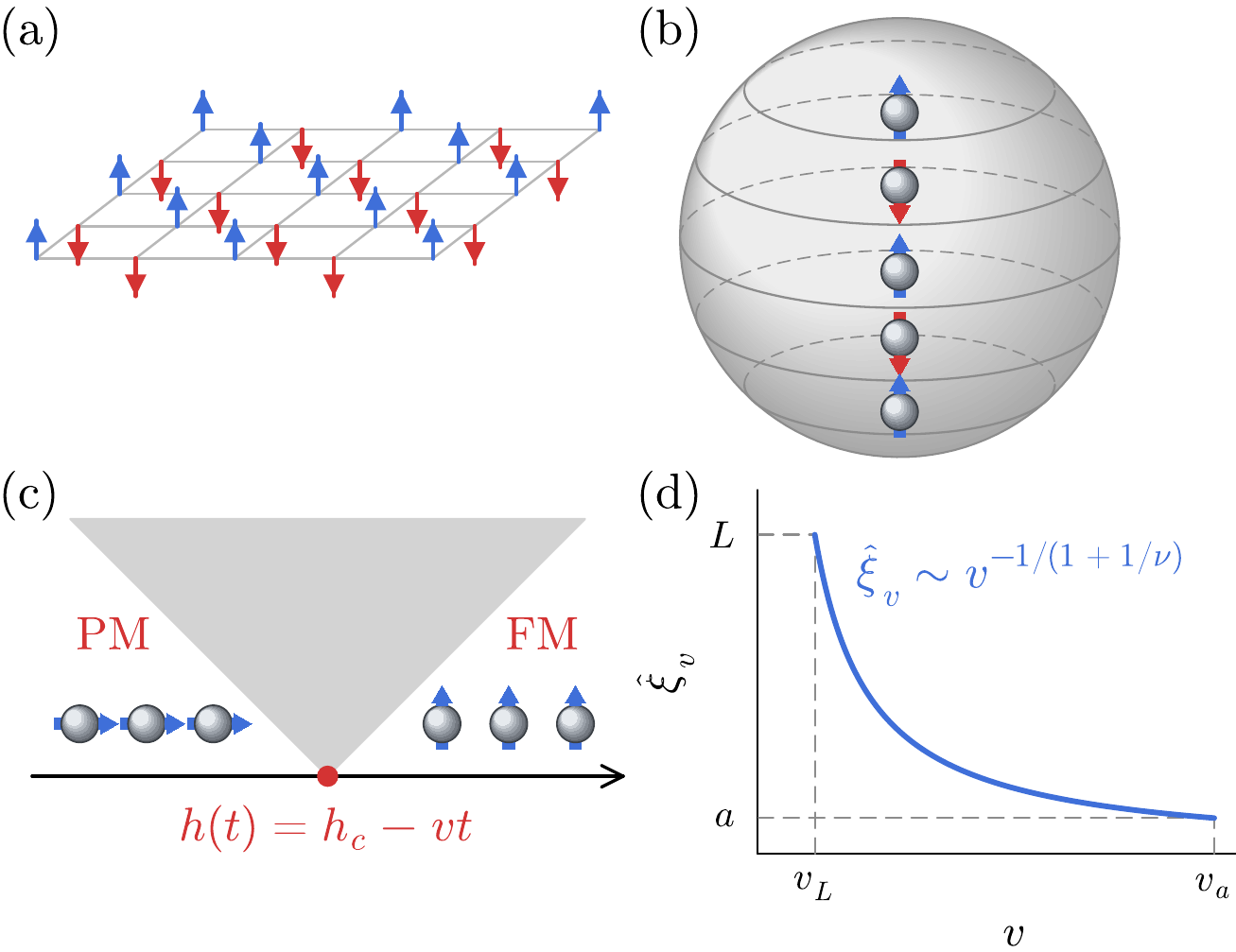}
    \caption{Schematic figures. (a) 2d Ising model on conventional square lattice. (b) Effective Ising model in the Landau orbital basis on the fuzzy sphere. (c) Linear quench protocol with rate $v$ for the transverse field across the critical point at $h_c$ from the paramagnetic (PM) phase to the ferromagnetic (FM) phase. (d) Dynamical correlation length $\hat{\xi}_v$ as a function of quench rate $v$, compared to linear system size $L$ and the short-distance cutoff $a$, with the corresponding rates labeled by $v_L$ and $v_a$ respectively.}
    \label{fig:schematic}
\end{figure}
The fuzzy sphere regularization of the Ising transition is essentially a quantum Hall ferromagnet-paramagnet transition \cite{zhu_uncovering_2023} on the sphere \cite{haldane-spehre-1985}. 
The original model of Ref.~\cite{zhu_uncovering_2023} is
\begin{equation}
\label{eq:hamiltonian}
    H=\int d\Omega_a d\Omega_b U_{ab}\left(\rho_a^0\rho_b^0-\rho_a^z\rho_b^z\right)-h\int d\Omega_a \rho_a^x,
\end{equation}
where $\rho_a^\mu\equiv \frac{1}{2}f^\dag(\Omega_a)\sigma^\mu f(\Omega_a)$ with $f\equiv (f_\up,f_\dn)^\mathrm{T}$ being the electron operators, and $\sigma^\mu$ with $\mu=0,x,y,z$ are Pauli matrices. $\Omega_{a/b}\equiv (\theta_{a/b},\phi_{a/b})$ denote the real-space locations on the sphere with $\theta,\phi$ being the polar and azimuthal angles respectively, and $U_{ab}\equiv U(\Omega_a,\Omega_b)$ is the interaction (in both density and spin channels) between electrons located at $\Omega_{a}$ and $\Omega_b$. The $2^\mathrm{nd}$ term represents the coupling to a transverse field in the $x$-direction.
With a $4\pi s$-monopole placed at the center of the sphere ($s$ can be half or whole integer), Landau levels form, with the lowest Landau level (LLL) orbitals given by the following wave functions
\begin{equation}
    \Phi_m(\theta,\phi)=\mathcal{N}_m \e^{im\phi}\cos^{s+m}\left( \frac{\theta}{2}\right)\sin^{s-m}\left( \frac{\theta}{2}\right),
\end{equation}
where $\mathcal{N}_m$ is a normalization factor with $m=-s,-s+1,...,s-1,s$ being the angular momentum quantum number, giving rise to a degeneracy of $2s+1$ on the LLL. 
The effective Ising transition becomes apparent when, at half-filling (with $N=2s+1$ electrons), $U_{ab}$ in Eq.~(\ref{eq:hamiltonian}) is chosen such that the density-density interaction gaps out the charge degrees of freedom by forming a charge quantum Hall insulator, and the spin-spin interaction becomes the ferromagnetic coupling. At low energy in the LLL, the physics is then dominated by the spin degrees of freedom, realizing the Ising transition when $h$ is tuned. Projection to the LLL can be done by restricting to the $(2s+1)$-degenerate LLL orbitals using 
\begin{equation}
    f_{\up/\dn}(\theta,\phi)=\sum_{m=-s}^{s}\Phi_m^*(\theta,\phi)c_{m,\up/\dn},
\end{equation}
where $c_{m,\up/\dn}$ is the electron operator in the $m$-th Landau orbital. The LLL-projected Hamiltonian is given by
\begin{equation}
\begin{aligned}
H_{LLL} = & \sum_{m_i} V_{m_1,m_2}^{m_3,m_4}\,
\delta_{m_1+m_3,m_2+m_4} \\
& \times \Bigl[
c_{m_1}^\dag c_{m_2} c_{m_3}^\dag c_{m_4} \\
& \qquad
- (c_{m_1}^\dag \sigma^z c_{m_2})
  (c_{m_3}^\dag \sigma^z c_{m_4})
\Bigr] \\
& - h \sum_m c_m^\dag \sigma^x c_m \,.
\end{aligned}
\end{equation}
where $c_m=(c_{m\up},c_{m\dn})^\mathrm{T}$ and $V_{m_1,m_2}^{m_3,m_4}$ can be expressed in terms of Haldane pseudo-potentials $V_l$ \cite{haldane-pseudo}. Here we follow the original paper \cite{zhu_uncovering_2023} and use the same ultra-local interaction $U_{ab}=g_0/R^2\delta(\Omega_{a}-\Omega_b)+g_1/R^4\nabla^2\delta(\Omega_{a}-\Omega_b)$, where $R$ is the radius of the sphere and $g_{0,1}$ are the coupling constants that can be absorbed into $V_l$. The eventual parameters used are $V_0=1,V_1=4.75$ with all other $V_l$ being zero, for which the Ising transition happens at $h_c\approx 3.16$ (see \cite{zhu_uncovering_2023} for more details).

Given the equilibrium critical point at transverse field $h_c$,  correlation length $\xi$ and  relaxation time $\tau$ scale 
\begin{equation}
\label{eq:scalings}
\xi\sim |h-h_c|^{-\nu}, \qquad
\tau\sim \xi^z \sim |h-h_c|^{-z\nu},
\end{equation}
where $\nu$ is the correlation length critical exponent and $z$ the dynamical critical exponent. For the 2D quantum Ising transition studied in this work, $z=1$ due to conformal symmetry and $\nu\approx 0.63$.  
We focus on the linear quench protocol with the time-dependent transverse field
\begin{equation}
h(t)=h_c-vt,
\end{equation}
where $v>0$ is the quench rate and $vt$ gives the effective distance in parameter space from criticality. As a result, the scaling relations given in Eq.~(\ref{eq:scalings}) also pick up time dependence: $\xi(t)\sim v^{-\nu}|t|^{-\nu},
\tau(t)\sim v^{-z\nu}|t|^{-z\nu}$. However, such time-dependent relations are only valid in the quasi-adiabatic regime where the time-evolved state remains the ground state of the time-dependent Hamiltonian, because the equilibrium relations in Eq.~(\ref{eq:scalings}) break down when excitations are created.

For a generic non-zero quench rate $v$ from an initial state far away from the critical point, i.e., with a large energy gap, the state evolves adiabatically at the beginning. When it becomes sufficiently close to criticality with a sufficiently small gap,  excitations are generated. In the KZ picture, the system enters the ``impulse regime'' where the relaxation time becomes large enough that the system becomes effectively frozen. The corresponding time scale, the freeze-out time $\hat{t}_v$, is defined such that $\xi(\hat{t}_v)\sim \hat{t}_v$, where the quench-rate dependence is made explicit. Making use of the scaling relations, we obtain 
\begin{equation}
\hat{t}_v\sim v^{-z\nu/(1+z\nu)}, \qquad
\hat{\xi}_v\equiv \xi(\hat{t}_v)\sim v^{-\nu/(1+z\nu)},
\label{eq:freeze-out-time-length}
\end{equation}
where $\hat{\xi}_v$ is the correlation length at the freeze-out time and it characterizes the typical size of the domains in symmetry-breaking transitions. As a consequence, the density of domain walls or topological defects follows the universal KZ scaling form~\cite{kibble1976topology,kibble1980some,zurek1985cosmological,zurek1996cosmological} $
n_\mathrm{defects} \sim \hat{\xi}_v^{-d} \sim v^{d\nu/(1+z\nu)}$, where $d$ is the spatial dimension. Other than the scaling of defect density, KZ physics also shows up in other observables, for instance, the squared magnetization, the correlation function and the excitation energy that we study in this work. 

In general, there are three different scaling regimes depending on the quench rate $v$ (Fig.~\ref{fig:schematic}(d)). (i) The adiabatic regime in the slow quench limit $v< v_L$, with $v_L$ defined using $\hat{\xi}_{v_L}\sim L$, such that in this regime $\hat{\xi}_v$ exceeds the system size $L$, hence the time evolution is adiabatic. (ii) The intermediate regime with $a<\hat{\xi}_v<L$, where $a$ is a short-distance cutoff. The system falls out of equilibrium with excitations generated during the quench.  KZ scaling emerges when $\hat{\xi}_v\ll L$ in this regime. (iii) The fast quench regime with $\hat{\xi}_v\lesssim a$. For a conventional lattice, $a$ is the lattice constant, but for the fuzzy sphere regularization of the Ising model in our work, a natural cutoff is the magnetic length. Given $N\sim L^2$ Landau orbitals, the magnetic length is approximately $L/\sqrt{N}\sim 1$. 

\paragraph{\textcolor{violet}{Numerical results}.}
The numerical calculation is done with the FuzzifiED package \cite{zhou_fuzzified_2025} adapted to TDVP \cite{tdvp} modified by the recently proposed bond expansion algorithm \cite{cbe} using ITensor \cite{fishman2022itensor}. We use $\delta t=0.05$ as the time step for the evolution, for which the truncation error due to finite time step is tested to be $\sim 0.1\%$ for $\langle m_z^2\rangle$ and $\lesssim 3\%$ for $Q$. The initial time $t_i$ is chosen with $|t_i|<\hat{t}_v$ for each quench rate $v$ to ensure that the evolution starts from the adiabatic regime. The maximum bond dimension used for TDVP is $D_\mathrm{max}=1000$, which is sufficient for the convergence of the observables. More details can be found in End Matter.

(i) \textbf{Squared magnetization.}
On the fuzzy sphere, the magnetization density operator is given by
\begin{equation}
    m_z=\frac{1}{N}\int d\Omega \ \rho^\mu(\Omega)=\frac{1}{N}\sum_{m=-s}^{s}\frac{1}{2}c_m^\dag\sigma^z c_m,
    \label{eq:m-definition}
\end{equation}
based on which  $\langle m_z^2\rangle$ can be readily calculated in both ED and TDVP simulations. When quenched to the critical point, if the physics is dominated by the correlation length $\hat{\xi}_v$, we have the following FTS form~\cite{Zhifangxu2005prb,Liu2014prb,huangyy2014prb},
\begin{equation}
    \label{eq:m-scaling}
    \langle m_z^2\rangle=\hat{\xi}_v^{-2\Delta}\mathcal{F}_m(L/\hat{\xi}_v),
\end{equation}
where $\mathcal{F}_m$ is the corresponding universal scaling function and $\Delta\approx 0.518$ is the scaling dimension of the order parameter field. 
\begin{figure}
    \centering
    \includegraphics[width=\linewidth]{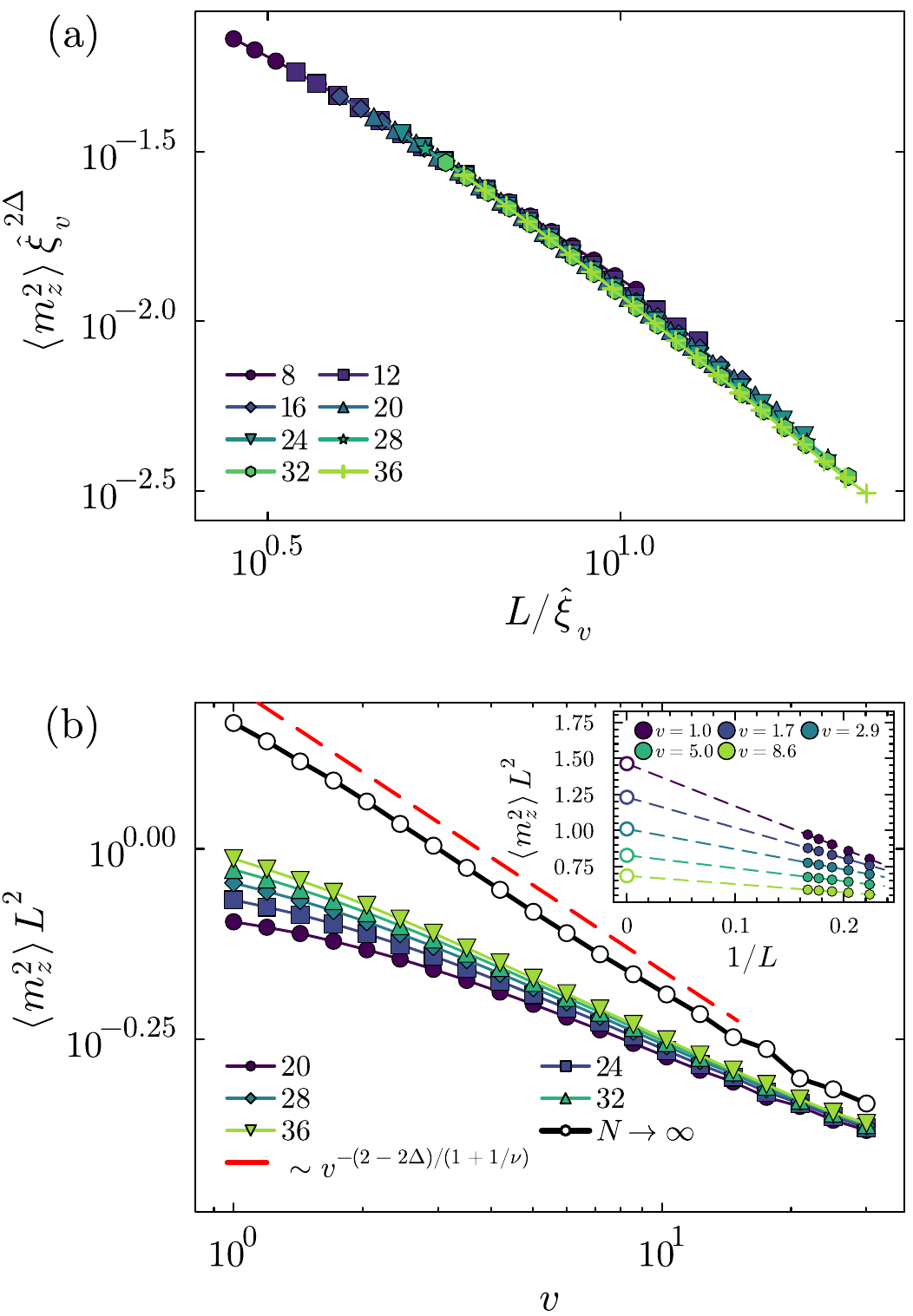}
    \caption{(a) Data collapse of $\langle m_z^2\rangle$ for system sizes ranging from $N=8$ to 36. The known critical exponents for the 3D Ising universality class are used for the plots. (b) Extrapolation of $\langle m_z^2\rangle L^2$ to the thermodynamic limit at each quench rate (black empty circle), with the inset showing the linear extrapolation at five representative quench rates. The red dashed line shows the KZ prediction \cite{liu_dynamic_2014} as a guide to the eye. The error bar from linear extrapolation is smaller than marker size. The error due to the finite time-step of the evolution is not directly reflected in the plot, but is estimated to be small enough that the conclusion is not changed qualitatively (see End Matter).}
    \label{fig:mag}
\end{figure}
Fig.~\ref{fig:mag}(a) shows the data collapse for system sizes ranging from $N=8$ to $36$ electrons. Fig.~\ref{fig:mag}(b) shows the extrapolation to the thermodynamic limit for $\langle m_z^2\rangle L^2$. For a decent range of quench rates, the extrapolated line follows the theoretical prediction for the KZ scaling indicated by the dashed line. The slope of the extrapolated line starts to differ from  KZ scaling for larger quench rates.  Such deviations indicate the crossover from the KZ scaling regime to the fast quench regime, where the values of $\langle m_z^2\rangle L^2$ for different system sizes start to converge. For the Ising model on a conventional lattice, $\langle m_z^2\rangle L^2$ eventually becomes system-size and quench-rate independent, giving rise to the $\langle m_z^2\rangle\sim L^{-2}$ \cite{liu_dynamic_2014}. This fast quench scaling is also verified in our case (see End Matter). In the End Matter, we also present the ED results for the slow quench limit, showing the expected adiabatic scaling $\langle m_z^2\rangle\sim L^{-2\Delta}$.

(ii) \textbf{Excitation energy density.}
\begin{figure}[t]
    \centering
    \includegraphics[width=\linewidth]{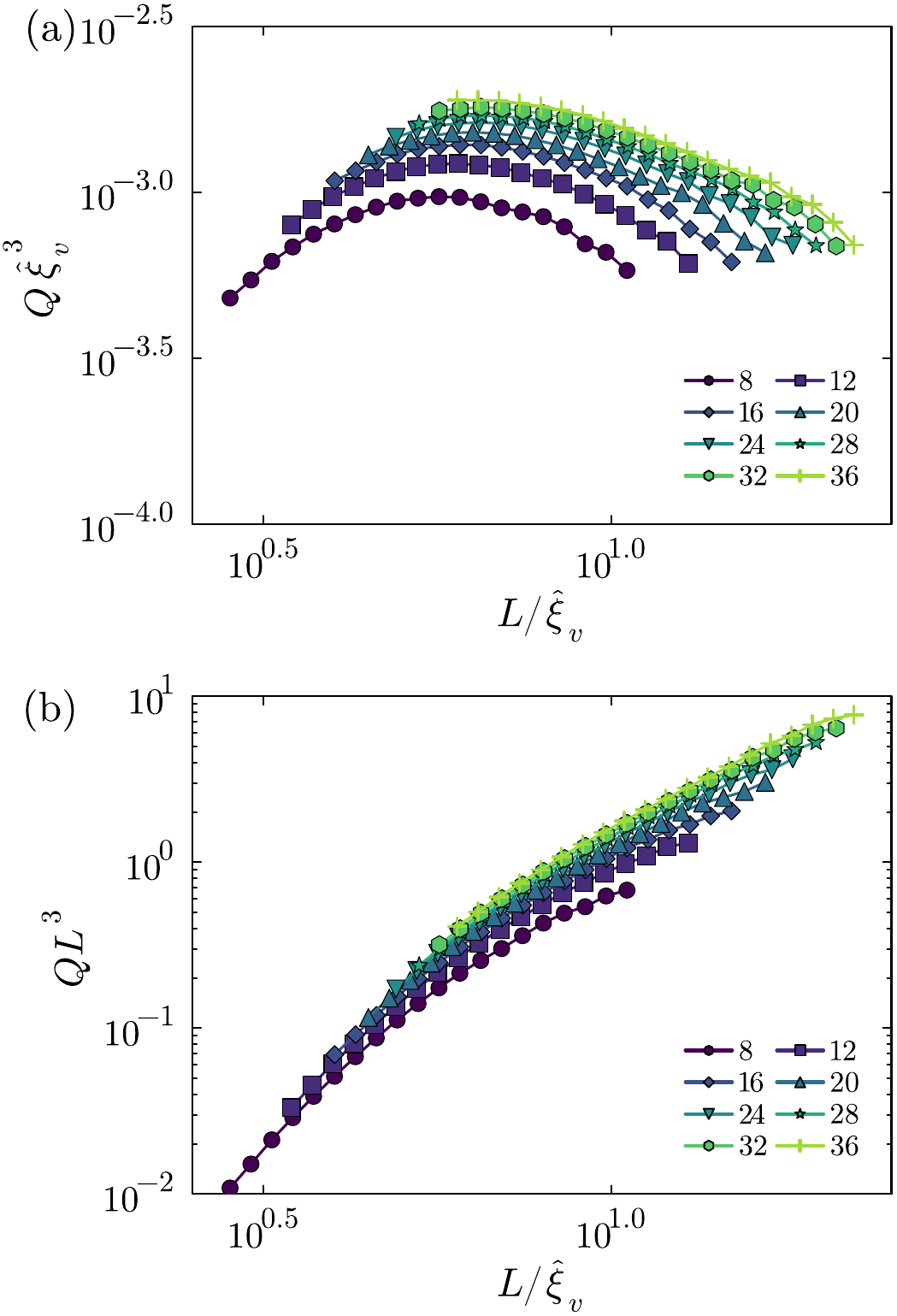}
    \caption{Excitation energy density $Q$, defined in Eq.~(\ref{eq:Q-definition}), measured at the critical point. (a) Scaling of $Q$ by $\hat{\xi}_v^3$ shows no satisfactory data collapse for the available sizes, in contrast to Fig.~\ref{fig:mag}(a). (b) Scaling of $Q$ by $L^3$ only shows signs of collapsing on the slow quench end.}
    \label{fig:Q}
\end{figure}
For the spherical geometry, it is difficult to directly measure the defects between the domains. Nevertheless, we can study instead a related quantity, the excitation energy density
\begin{equation}
\label{eq:Q-definition}
    Q(t)\equiv \frac{1}{N}\left(\langle \psi(t)|H(t)|\psi(t)\rangle-E_0(t)\right),
\end{equation}
where $\ket{\psi(t)}$ is the time-evolved state and $E_0(t)$ is the instantaneous ground state energy of $H(t)$. For unitary quantum dynamics, a non-zero $Q$ implies that the time-evolved state deviates from the instantaneous ground state and contains states from the excited part of the spectrum, which correspond to the formation of defects. Similar to Eq.~(\ref{eq:m-scaling}), $Q$ at $t=0$ follows
\begin{equation}
    \label{eq:Q-scaling} Q=\hat{\xi}_v^{-3}\mathcal{F}_Q(L/\hat{\xi}_v),
\end{equation}
because $Q$ has dimension $d+z=3$. $\mathcal{F}_Q$ is the corresponding universal scaling function. Fig.~\ref{fig:Q}(a) shows the attempt to collapse the data based on Eq.~(\ref{eq:Q-scaling}). However, in contrast to Fig.~\ref{fig:mag}(a), no satisfactory data collapse is achieved for the system sizes used. As a comparison, Fig.~\ref{fig:Q}(b) with $Q$ scaled by $L^3$ only shows signs of collapse on the slow quench side, which is consistent with the ED results for $Q$ in the slow quench limit demonstrated in End Matter. The failure of data collapse indicates that the scaling limit for $Q$ has not been reached with the available system sizes.
This is likely due to the symmetries in the spherical setup, which is preserved by the initial state, the Hamiltonian, and the quench protocol. As a result, the possible excited states in the non-equilibrium process are also constrained to the same symmetry sector a sthe ground state, which in our case is the $l=0$ angular momentum sector of the $SO(3)$, the $+1$ sector of the particle-hole symmetry (spacetime parity of CFT), the $+1$ sector of the Ising $\mathbb{Z}_2$ and the $+1$ sector of the $\pi$-rotation around the $y$-axis \cite{zhu_uncovering_2023}. The symmetry allowed spectrum becomes comparatively sparse, making the effective finite-size gap at the critical point significantly larger, which in turn requires  larger system sizes to achieve the scaling regime (see End Matter for the numerical evidence of symmetry-enforced level sparsity). It is plausible that non-energy quantities like $\langle m_z^2\rangle$ and correlation function (see below) are less affected in this respect. In the End Matter, we also show the adiabatic scaling $Q\sim v^2$ in the slow quench limit using ED for small systems, agreeing with adiabatic perturbation theory prediction \cite{de_grandi_quench_2010}.

\begin{figure}[t]
    \centering
    \includegraphics[width=\linewidth]{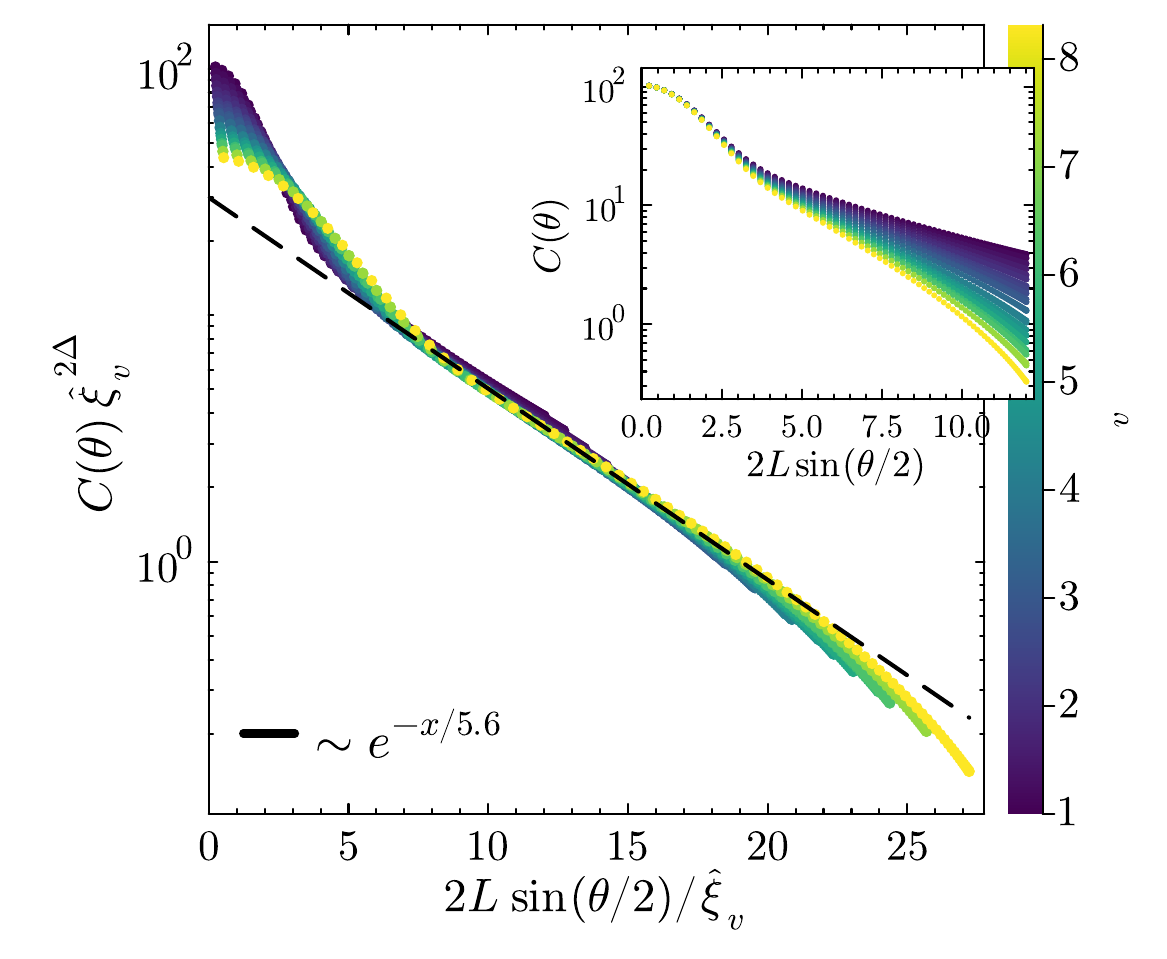}
    \caption{Two-point correlation function on the sphere for $N=36$ electrons across a wide range of quench rates (chosen such that they are approximately equally spaced on logarithmic scale). Data collapse is done based on Eq.~(\ref{eq:corr-scaling}). The collapsed part follows exponential decay indicated by the dashed line. Inset shows the raw data.}
    \label{fig:corr}
\end{figure}

(iii) \textbf{Two-point correlation function.}
At the equilibrium critical point, the equal-time two-point correlation function for the order parameter $m_z$ on the fuzzy sphere with radius $R\sim \sqrt{N}\sim L$ can be obtained through a Weyl transformation from the correlation function in flat Euclidean space \cite{zhou_fuzzified_2025,han-four-point-2025}, and is given by 
\begin{equation}
   C_{\mathrm{eq}}(\theta)\equiv \langle m_z(\theta)m_z(0)\rangle\sim\frac{1}{(2L\sin(\theta/2))^{2\Delta}},
\end{equation}
where $\theta$ is the polar angle and $2L\sin(\theta/2)$ is the effective distance on the sphere. In the non-equilibrium case, when quenched to the critical point, we have the following scaling form for the correlation function,
\begin{equation}
    \label{eq:corr-scaling}
C(\theta)=\hat{\xi}_v^{-2\Delta}\mathcal{F}_C(L\sin(\theta/2)/\hat{\xi}_v),
\end{equation}
where $\mathcal{F}_C$ is a universal scaling function for the two-point correlation.
Fig.~\ref{fig:corr} shows the correlation function for a wide range of quench rates for $N=36$ electrons. When scaled properly based on Eq.~(\ref{eq:corr-scaling}), the data points collapse nicely to a universal curve. Deviations from the universal curve occur at both the  short- and the long-distance end, where  finite-size effects are  most pronounced \cite{han-four-point-2025}. The collapsed part can be approximately fitted by an exponential function $\sim e^{-x/5.6}$ (dashed line), which indicates the relation $\hat{\xi}_v\approx 5.6v^{-1/(1+1/\nu)}$, because in the plots we have being using $\hat{\xi}_v=v^{-1/(1+1/\nu)}$, i.e. a unit coefficient in the scaling relation Eq.~(\ref{eq:freeze-out-time-length}). This estimate is also used to locate the boundary between adiabatic regime and impulse regime for each quench rate $v$, based on which the initial state can be reliably chosen such that the time evolution starts from the adiabatic regime. 

\paragraph{\textcolor{violet}{Discussion and outlook}.}
In conclusion, we have demonstrated that the fuzzy sphere regularization provides a powerful framework for studying non-equilibrium quantum critical dynamics in two dimensions, enabling a quantitative test of Kibble-Zurek mechanism and the full finite-time scaling in the transverse-field Ising model. By overcoming key limitations of conventional lattice approaches for real-time simulations in higher dimensions, our work establishes a new route toward precision studies of universal dynamics in interacting quantum systems. Looking forward, this framework can be naturally extended to many other two-dimensional quantum critical points \cite{fuzzy-review}, including models with continuous symmetries, topological phase transitions, and deconfined criticality, opening the door to systematic investigations of non-equilibrium universality across a broad class of strongly correlated quantum matter.

\textbf{Code Availability}. The main code used to do the TDVP simulation in this work is publicly available at \url{https://github.com/mengzeng-phys/2d-ising-kibble-zurek-fuzzy-sphere}.

\section*{Acknowledgment}
MZ would like to thank Yi-Zhuang You, Runze Chi, Lei Su, Dam Thanh Son, Andrea Cappelli, Zohar Komargodski, Gianluca Teza, Federico Balducci, Johannes Hofmann, Shuai A Chen, Masudul Haque for discussions. MZ is supported in part by the Alexander von Humboldt Foundation through a Humboldt Research Fellowship. SY is supported by the National Natural Science Foundation of China (Grants No. 12222515), the Research Center for Magnetoelectric Physics of Guangdong Province (Grant No. 2024B0303390001), the Guangdong Provincial Key Laboratory of Magnetoelectric Physics and Devices (Grant No. 2022B1212010008), the Science and Technology Projects in Guangzhou City (Grant No. 2025A04J5408), and Quantum Science and Technology-National Science and Technology Major Project (Grant No.2025ZD0300400). This work was in part supported by the Deutsche Forschungsgemeinschaft under grants FOR 5522 (project-id 499180199) and the cluster of excellence ctd.qmat (EXC 2147, project-id 390858490).

\bibliography{main.bib} 

@article{kibble1976topology,
  title={Topology of cosmic domains and strings},
  author={Kibble, Thomas WB},
  journal={Journal of Physics A: Mathematical and General},
  volume={9},
  number={8},
  pages={1387--1398},
  year={1976},
  url={https://iopscience.iop.org/article/10.1088/0305-4470/9/8/029}
}

@article{kibble1980some,
  title={Some implications of a cosmological phase transition},
  author={Kibble, Tom WB},
  journal={Physics Reports},
  volume={67},
  number={1},
  pages={183--199},
  year={1980},
  publisher={Elsevier},
  url={https://www.sciencedirect.com/science/article/abs/pii/0370157380900915}
}

@article{zurek1985cosmological,
  title={Cosmological experiments in superfluid helium?},
  author={Zurek, Wojciech H},
  journal={Nature},
  volume={317},
  number={6037},
  pages={505--508},
  year={1985},
  publisher={Nature Publishing Group UK London},
  url={https://www.nature.com/articles/317505a0}
}

@article{zurek1996cosmological,
  title={Cosmological experiments in condensed matter systems},
  author={Zurek, Wojciech Hubert},
  journal={Physics Reports},
  volume={276},
  number={4},
  pages={177--221},
  year={1996},
  publisher={Elsevier},
  url={https://www.sciencedirect.com/science/article/abs/pii/S0370157396000099?via%3Dihub}
}

@article{shu_equilibration_2025,
	title = {Equilibration of topological defects near the deconfined quantum multicritical point},
	volume = {16},
	copyright = {2025 The Author(s)},
	issn = {2041-1723},
	url = {https://www.nature.com/articles/s41467-025-58477-z},
	doi = {10.1038/s41467-025-58477-z},
	number = {1},
	urldate = {2025-07-30},
	journal = {Nature Communications},
	publisher = {Nature Publishing Group},
	author = {Shu, Yu-Rong and Jian, Shao-Kai and Sandvik, Anders W. and Yin, Shuai},
	month = apr,
	year = {2025},
	pages = {3402}
}

@article{damski_simplest_2005,
	title = {The {Simplest} {Quantum} {Model} {Supporting} the {Kibble}-{Zurek} {Mechanism} of {Topological} {Defect} {Production}: {Landau}-{Zener} {Transitions} from a {New} {Perspective}},
	volume = {95},
	shorttitle = {The {Simplest} {Quantum} {Model} {Supporting} the {Kibble}-{Zurek} {Mechanism} of {Topological} {Defect} {Production}},
	url = {https://link.aps.org/doi/10.1103/PhysRevLett.95.035701},
	doi = {10.1103/PhysRevLett.95.035701},
	number = {3},
	urldate = {2025-07-31},
	journal = {Physical Review Letters},
	publisher = {American Physical Society},
	author = {Damski, Bogdan},
	month = jul,
	year = {2005},
	pages = {035701}
}

@article{lee-yang-2017,
	title = {Kibble-{Zurek} {Scaling} in the {Yang}-{Lee} {Edge} {Singularity}},
	volume = {118},
	url = {https://link.aps.org/doi/10.1103/PhysRevLett.118.065701},
	doi = {10.1103/PhysRevLett.118.065701},
	number = {6},
	urldate = {2025-08-04},
	journal = {Physical Review Letters},
	publisher = {American Physical Society},
	author = {Yin, Shuai and Huang, Guang-Yao and Lo, Chung-Yu and Chen, Pochung},
	month = feb,
	year = {2017},
	pages = {065701}
}

@article{dora_kibble-zurek_2019,
	title = {The {Kibble}-{Zurek} mechanism at exceptional points},
	volume = {10},
	copyright = {2019 The Author(s)},
	issn = {2041-1723},
	url = {https://www.nature.com/articles/s41467-019-10048-9},
	doi = {10.1038/s41467-019-10048-9},
	number = {1},
	urldate = {2025-08-05},
	journal = {Nature Communications},
	publisher = {Nature Publishing Group},
	author = {Dora, BalÃ¡zs and Heyl, Markus and Moessner, Roderich},
	month = may,
	year = {2019},
	pages = {2254}
}

@article{polkovnikov_colloquium_2011,
	title = {Colloquium: {Nonequilibrium} dynamics of closed interacting quantum systems},
	volume = {83},
	shorttitle = {Colloquium},
	url = {https://link.aps.org/doi/10.1103/RevModPhys.83.863},
	doi = {10.1103/RevModPhys.83.863},
	number = {3},
	urldate = {2025-08-05},
	journal = {Reviews of Modern Physics},
	publisher = {American Physical Society},
	author = {Polkovnikov, Anatoli and Sengupta, Krishnendu and Silva, Alessandro and Vengalattore, Mukund},
	month = aug,
	year = {2011},
	pages = {863--883},
}

@article{sengupta_exact_2008,
	title = {Exact {Results} for {Quench} {Dynamics} and {Defect} {Production} in a {Two}-{Dimensional} {Model}},
	volume = {100},
	url = {https://link.aps.org/doi/10.1103/PhysRevLett.100.077204},
	doi = {10.1103/PhysRevLett.100.077204},
	number = {7},
	urldate = {2025-08-05},
	journal = {Physical Review Letters},
	publisher = {American Physical Society},
	author = {Sengupta, K. and Sen, Diptiman and Mondal, Shreyoshi},
	month = feb,
	year = {2008},
	pages = {077204},
}

@article{chandran_kibble-zurek_2012,
	title = {Kibble-{Zurek} problem: {Universality} and the scaling limit},
	volume = {86},
	shorttitle = {Kibble-{Zurek} problem},
	url = {https://link.aps.org/doi/10.1103/PhysRevB.86.064304},
	doi = {10.1103/PhysRevB.86.064304},
	number = {6},
	urldate = {2025-08-22},
	journal = {Physical Review B},
	publisher = {American Physical Society},
	author = {Chandran, Anushya and Erez, Amir and Gubser, Steven S. and Sondhi, S. L.},
	month = aug,
	year = {2012},
	pages = {064304},
}

@article{kolodrubetz_nonequilibrium_2012,
	title = {Nonequilibrium {Dynamic} {Critical} {Scaling} of the {Quantum} {Ising} {Chain}},
	volume = {109},
	url = {https://link.aps.org/doi/10.1103/PhysRevLett.109.015701},
	doi = {10.1103/PhysRevLett.109.015701},
	number = {1},
	urldate = {2025-08-22},
	journal = {Physical Review Letters},
	publisher = {American Physical Society},
	author = {Kolodrubetz, Michael and Clark, Bryan K. and Huse, David A.},
	month = jul,
	year = {2012},
	pages = {015701},
}

@article{dziarmaga_dynamics_2005,
	title = {Dynamics of a {Quantum} {Phase} {Transition}: {Exact} {Solution} of the {Quantum} {Ising} {Model}},
	volume = {95},
	shorttitle = {Dynamics of a {Quantum} {Phase} {Transition}},
	url = {https://link.aps.org/doi/10.1103/PhysRevLett.95.245701},
	doi = {10.1103/PhysRevLett.95.245701},
	number = {24},
	urldate = {2025-08-22},
	journal = {Physical Review Letters},
	publisher = {American Physical Society},
	author = {Dziarmaga, Jacek},
	month = dec,
	year = {2005},
	pages = {245701},
}

@article{zeng_universal_2023,
	title = {Universal {Breakdown} of {Kibble}-{Zurek} {Scaling} in {Fast} {Quenches} across a {Phase} {Transition}},
	volume = {130},
	issn = {0031-9007, 1079-7114},
	url = {https://link.aps.org/doi/10.1103/PhysRevLett.130.060402},
	doi = {10.1103/PhysRevLett.130.060402},
	number = {6},
	urldate = {2025-10-21},
	journal = {Physical Review Letters},
	author = {Zeng, Hua-Bi and Xia, Chuan-Yin and Del Campo, Adolfo},
	month = feb,
	year = {2023},
	pages = {060402},
}

@article{schmitt_quantum_2022,
	title = {Quantum phase transition dynamics in the two-dimensional transverse-field {Ising} model},
	volume = {8},
	url = {https://www.science.org/doi/10.1126/sciadv.abl6850},
	doi = {10.1126/sciadv.abl6850},
	number = {37},
	urldate = {2025-11-17},
	journal = {Science Advances},
	publisher = {American Association for the Advancement of Science},
	author = {Schmitt, Markus and Rams, Marek M. and Dziarmaga, Jacek and Heyl, Markus and Zurek, Wojciech H.},
	month = sep,
	year = {2022},
	pages = {eabl6850},
}

@article{keesling_quantum_2019,
	title = {Quantum {Kibble}-{Zurek} mechanism and critical dynamics on a programmable {Rydberg} simulator},
	volume = {568},
	copyright = {2019 The Author(s), under exclusive licence to Springer Nature Limited},
	issn = {1476-4687},
	url = {https://www.nature.com/articles/s41586-019-1070-1},
	doi = {10.1038/s41586-019-1070-1},
	number = {7751},
	urldate = {2025-11-17},
	journal = {Nature},
	publisher = {Nature Publishing Group},
	author = {Keesling, Alexander and Omran, Ahmed and Levine, Harry and Bernien, Hannes and Pichler, Hannes and Choi, Soonwon and Samajdar, Rhine and Schwartz, Sylvain and Silvi, Pietro and Sachdev, Subir and Zoller, Peter and Endres, Manuel and Greiner, Markus and Vuletic, Vladan and Lukin, Mikhail D.},
	month = apr,
	year = {2019},
	pages = {207--211}
}

@article{de_grandi_quench_2010,
	title = {Quench dynamics near a quantum critical point},
	volume = {81},
	copyright = {http://link.aps.org/licenses/aps-default-license},
	issn = {1098-0121, 1550-235X},
	url = {https://link.aps.org/doi/10.1103/PhysRevB.81.012303},
	doi = {10.1103/PhysRevB.81.012303},
	number = {1},
	urldate = {2026-02-05},
	journal = {Physical Review B},
	author = {De Grandi, C. and Gritsev, V. and Polkovnikov, A.},
	month = jan,
	year = {2010},
	pages = {012303},
}

@article{liu_dynamic_2014,
	title = {Dynamic scaling at classical phase transitions approached through nonequilibrium quenching},
	volume = {89},
	copyright = {http://link.aps.org/licenses/aps-default-license},
	issn = {1098-0121, 1550-235X},
	url = {https://link.aps.org/doi/10.1103/PhysRevB.89.054307},
	doi = {10.1103/PhysRevB.89.054307},
	number = {5},
	urldate = {2026-02-06},
	journal = {Physical Review B},
	author = {Liu, Cheng-Wei and Polkovnikov, Anatoli and Sandvik, Anders W.},
	month = feb,
	year = {2014},
	pages = {054307}
}

@article{zhu_uncovering_2023,
	title = {Uncovering {Conformal} {Symmetry} in the {3D} {Ising} {Transition}: {State}-{Operator} {Correspondence} from a {Quantum} {Fuzzy} {Sphere} {Regularization}},
	volume = {13},
	shorttitle = {Uncovering {Conformal} {Symmetry} in the {3D} {Ising} {Transition}},
	url = {https://link.aps.org/doi/10.1103/PhysRevX.13.021009},
	doi = {10.1103/PhysRevX.13.021009},
	number = {2},
	urldate = {2025-07-02},
	journal = {Physical Review X},
	publisher = {American Physical Society},
	author = {Zhu, Wei and Han, Chao and Huffman, Emilie and Hofmann, Johannes S. and He, Yin-Chen},
	month = apr,
	year = {2023},
	pages = {021009},
}

@article{zhou_so5_2024,
	title = {{SO}(5) {Deconfined} {Phase} {Transition} under the {Fuzzy}-{Sphere} {Microscope}: {Approximate} {Conformal} {Symmetry}, {Pseudo}-{Criticality}, and {Operator} {Spectrum}},
	volume = {14},
	shorttitle = {{SO}(5) {Deconfined} {Phase} {Transition} under the {Fuzzy}-{Sphere} {Microscope}},
	url = {https://link.aps.org/doi/10.1103/PhysRevX.14.021044},
	doi = {10.1103/PhysRevX.14.021044},
	number = {2},
	urldate = {2025-07-02},
	journal = {Physical Review X},
	publisher = {American Physical Society},
	author = {Zhou, Zheng and Hu, Liangdong and Zhu, W. and He, Yin-Chen},
	month = jun,
	year = {2024},
	pages = {021044},
}

@article{fuzzy-review,
  title={{A Fuzzy Sphere Journey in Critical Phenomena}},
  author={He, Yin-Chen and Zhu, W},
  journal={Annual Review of Condensed Matter Physics},
  url = {https://www.annualreviews.org/content/journals/10.1146/annurev-conmatphys-031424-020256#abstract_content},
  volume={17},
  number={1},
  pages={1--25},
  year={2026},
  publisher={Annual Reviews}
}

@article{hu_solving_2024,
	title = {Solving conformal defects in {3D} conformal field theory using fuzzy sphere regularization},
	volume = {15},
	copyright = {2024 The Author(s)},
	issn = {2041-1723},
	url = {https://www.nature.com/articles/s41467-024-47978-y},
	doi = {10.1038/s41467-024-47978-y},
	number = {1},
	urldate = {2025-07-30},
	journal = {Nature Communications},
	publisher = {Nature Publishing Group},
	author = {Hu, Liangdong and He, Yin-Chen and Zhu, W.},
	month = apr,
	year = {2024},
	pages = {3659},
}

@article{zhou_fuzzified_2025,
  title={{FuzzifiED: Julia package for numerics on the fuzzy sphere}},
  author={Zhou, Zheng},
  journal={arXiv:2503.00100},
  year={2025},
  url = {http://arxiv.org/abs/2503.00100},
}

@article{fan2025simulating-lee-yang,
  title={{Simulating the non-unitary Yang-Lee conformal field theory on the fuzzy sphere}},
  author={Fan, Ruihua and Dong, Junkai and Vishwanath, Ashvin},
  journal={arXiv:2505.06342},
  year={2025},
  url = {http://arxiv.org/abs/2505.06342}
}

@article{potts-2025,
  title={{Microscopic study of the three-dimensional Potts phase transition via fuzzy sphere regularization}},
  author={Yang, Shuai and Yue, Yan-Guang and Tang, Yin and Han, Chao and Zhu, W and Chen, Yan},
  journal={Physical Review B},
  volume={112},
  number={2},
  pages={024436},
  year={2025},
  publisher={APS},
  url={https://journals.aps.org/prb/abstract/10.1103/x1qn-x6xb}
}

@article{han_conformal_2024,
	title = {{Conformal operator content of the {Wilson}-{Fisher} transition on fuzzy sphere bilayers}},
	volume = {110},
	url = {https://link.aps.org/doi/10.1103/PhysRevB.110.115113},
	doi = {10.1103/PhysRevB.110.115113},
	number = {11},
	urldate = {2025-08-27},
	journal = {Physical Review B},
	publisher = {American Physical Society},
	author = {Han, Chao and Hu, Liangdong and Zhu, W.},
	month = sep,
	year = {2024},
	pages = {115113},
}

@article{fishman2022itensor,
  title={{The ITensor software library for tensor network calculations}},
  author={Fishman, Matthew and White, Steven and Stoudenmire, Edwin Miles},
  journal={SciPost Physics Codebases},
  pages={004},
  year={2022},
  url ={https://www.scipost.org/SciPostPhysCodeb.4?acad_field_slug=politicalscience}
}

@article{tdvp,
  title = {{Time-Dependent Variational Principle for Quantum Lattices}},
  author = {Haegeman, Jutho and Cirac, J. Ignacio and Osborne, Tobias J. and Pi\ifmmode \check{z}\else \v{z}\fi{}orn, Iztok and Verschelde, Henri and Verstraete, Frank},
  journal = {Phys. Rev. Lett.},
  volume = {107},
  issue = {7},
  pages = {070601},
  numpages = {5},
  year = {2011},
  month = {Aug},
  publisher = {American Physical Society},
  doi = {10.1103/PhysRevLett.107.070601},
  url = {https://link.aps.org/doi/10.1103/PhysRevLett.107.070601}
}

@article{cbe,
  title = {{Time-Dependent Variational Principle with Controlled Bond Expansion for Matrix Product States}},
  author = {Li, Jheng-Wei and Gleis, Andreas and von Delft, Jan},
  journal = {Phys. Rev. Lett.},
  volume = {133},
  issue = {2},
  pages = {026401},
  numpages = {9},
  year = {2024},
  month = {Jul},
  publisher = {American Physical Society},
  doi = {10.1103/PhysRevLett.133.026401},
  url = {https://link.aps.org/doi/10.1103/PhysRevLett.133.026401}
}

@article{beyond-KZ-roderich,
  title = {{Dynamics and correlations at a quantum phase transition beyond Kibble-Zurek}},
  author = {Roychowdhury, Krishanu and Moessner, Roderich and Das, Arnab},
  journal = {Phys. Rev. B},
  volume = {104},
  issue = {1},
  pages = {014406},
  numpages = {11},
  year = {2021},
  month = {Jul},
  publisher = {American Physical Society},
  doi = {10.1103/PhysRevB.104.014406},
  url = {https://link.aps.org/doi/10.1103/PhysRevB.104.014406}
}

@article{spacetime-2016,
  title = {{Space and time renormalization in phase transition dynamics}},
  author = {Francuz, Anna and Dziarmaga, Jacek and Gardas, Bart\l{}omiej and Zurek, Wojciech H.},
  journal = {Phys. Rev. B},
  volume = {93},
  issue = {7},
  pages = {075134},
  numpages = {14},
  year = {2016},
  month = {Feb},
  publisher = {American Physical Society},
  doi = {10.1103/PhysRevB.93.075134},
  url = {https://link.aps.org/doi/10.1103/PhysRevB.93.075134}
}

@article{masud-2018,
  title={{Dynamics and level statistics of interacting fermions in the lowest Landau level}},
  author={Fremling, M and Repellin, C{\'e}cile and St{\'e}phan, Jean-Marie and Moran, N and Slingerland, JK and Haque, Masudul},
  journal={New Journal of Physics},
  volume={20},
  number={10},
  pages={103036},
  year={2018},
  publisher={IOP Publishing},
  url={https://iopscience.iop.org/article/10.1088/1367-2630/aae73f}
}

@article{han-four-point-2025,
  title = {{Conformal four-point correlators of the three-dimensional Ising transition via the quantum fuzzy sphere}},
  author = {Han, Chao and Hu, Liangdong and Zhu, W. and He, Yin-Chen},
  journal = {Phys. Rev. B},
  volume = {108},
  issue = {23},
  pages = {235123},
  numpages = {7},
  year = {2023},
  month = {Dec},
  publisher = {American Physical Society},
  doi = {10.1103/PhysRevB.108.235123},
  url = {https://link.aps.org/doi/10.1103/PhysRevB.108.235123}
}

@article{review-heyl2018,
  title={{Dynamical quantum phase transitions: a review}},
  author={Heyl, Markus},
  journal={Reports on Progress in Physics},
  volume={81},
  number={5},
  pages={054001},
  year={2018},
  publisher={IOP Publishing},
  url={https://iopscience.iop.org/article/10.1088/1361-6633/aaaf9a/meta}
}

@article{heyl-dynamic-ising-transition-2013,
  title = {{Dynamical Quantum Phase Transitions in the Transverse-Field Ising Model}},
  author = {Heyl, M. and Polkovnikov, A. and Kehrein, S.},
  journal = {Phys. Rev. Lett.},
  volume = {110},
  issue = {13},
  pages = {135704},
  numpages = {5},
  year = {2013},
  month = {Mar},
  publisher = {American Physical Society},
  doi = {10.1103/PhysRevLett.110.135704},
  url = {https://link.aps.org/doi/10.1103/PhysRevLett.110.135704}
}

@article{heyl-broken-symmetry-2014,
  title = {{Dynamical Quantum Phase Transitions in Systems with Broken-Symmetry Phases}},
  author = {Heyl, M.},
  journal = {Phys. Rev. Lett.},
  volume = {113},
  issue = {20},
  pages = {205701},
  numpages = {5},
  year = {2014},
  month = {Nov},
  publisher = {American Physical Society},
  doi = {10.1103/PhysRevLett.113.205701},
  url = {https://link.aps.org/doi/10.1103/PhysRevLett.113.205701}
}

@article{heyl-scaling-universality-2015,
  title = {{Scaling and Universality at Dynamical Quantum Phase Transitions}},
  author = {Heyl, Markus},
  journal = {Phys. Rev. Lett.},
  volume = {115},
  issue = {14},
  pages = {140602},
  numpages = {5},
  year = {2015},
  month = {Oct},
  publisher = {American Physical Society},
  doi = {10.1103/PhysRevLett.115.140602},
  url = {https://link.aps.org/doi/10.1103/PhysRevLett.115.140602}
}

@article{heyl-observation-2017,
  title = {{Direct Observation of Dynamical Quantum Phase Transitions in an Interacting Many-Body System}},
  author = {Jurcevic, P. and Shen, H. and Hauke, P. and Maier, C. and Brydges, T. and Hempel, C. and Lanyon, B. P. and Heyl, M. and Blatt, R. and Roos, C. F.},
  journal = {Phys. Rev. Lett.},
  volume = {119},
  issue = {8},
  pages = {080501},
  numpages = {5},
  year = {2017},
  month = {Aug},
  publisher = {American Physical Society},
  doi = {10.1103/PhysRevLett.119.080501},
  url = {https://link.aps.org/doi/10.1103/PhysRevLett.119.080501}
}

@article{markus-2d-neuralnet,
  title = {{Quantum Many-Body Dynamics in Two Dimensions with Artificial Neural Networks}},
  author = {Schmitt, Markus and Heyl, Markus},
  journal = {Phys. Rev. Lett.},
  volume = {125},
  issue = {10},
  pages = {100503},
  numpages = {7},
  year = {2020},
  month = {Sep},
  publisher = {American Physical Society},
  doi = {10.1103/PhysRevLett.125.100503},
  url = {https://link.aps.org/doi/10.1103/PhysRevLett.125.100503}
}

@article{3d-ising-2026,
  title={{Real-time Dynamics in 3D for up to 1000 Qubits with Neural Quantum States: Quenches and the Quantum Kibble--Zurek Mechanism}},
  author={Naik, Vighnesh Dattatraya and Sun, Zheng-Hang and Heyl, Markus},
  journal={arXiv:2604.05032},
  year={2026},
  url={https://arxiv.org/abs/2604.05032}
}

@article{federico-2025,
  title = {{Driving a quantum phase transition at an arbitrary rate: Exact solution of the transverse-field Ising model}},
  author = {Grabarits, Andr\'as and Balducci, Federico and del Campo, Adolfo},
  journal = {Phys. Rev. A},
  volume = {111},
  issue = {4},
  pages = {042207},
  numpages = {17},
  year = {2025},
  month = {Apr},
  publisher = {American Physical Society},
  doi = {10.1103/PhysRevA.111.042207},
  url = {https://link.aps.org/doi/10.1103/PhysRevA.111.042207}
}

@article{braun-emergence-coherence-2015,
author = {Simon Braun  and Mathis Friesdorf  and Sean S. Hodgman  and Michael Schreiber  and Jens Philipp Ronzheimer  and Arnau Riera  and Marco del Rey  and Immanuel Bloch  and Jens Eisert  and Ulrich Schneider },
title = {{Emergence of coherence and the dynamics of quantum phase transitions}},
journal = {Proceedings of the National Academy of Sciences},
volume = {112},
number = {12},
pages = {3641-3646},
year = {2015},
doi = {10.1073/pnas.1408861112},
URL = {https://www.pnas.org/doi/abs/10.1073/pnas.1408861112}
}

@article{du2023kibble,
  title={{Kibble--Zurek mechanism of Ising domains}},
  author={Du, Kai and Fang, Xiaochen and Won, Choongjae and De, Chandan and Huang, Fei-Ting and Xu, Wenqian and You, Hoydoo and G{\'o}mez-Ruiz, Fernando J and Del Campo, Adolfo and Cheong, Sang-Wook},
  journal={Nature Physics},
  volume={19},
  number={10},
  pages={1495--1501},
  year={2023},
  publisher={Nature Publishing Group UK London},
  url={https://www.nature.com/articles/s41567-023-02112-5}
}

@article{weinberg2025defects,
  title={{Defects and their Time Scales in Quantum and Classical Annealing of the Two-Dimensional Ising Model}},
  author={Weinberg, Phillip and Xu, Na and Sandvik, Anders W},
  journal={arXiv:2507.09273},
  year={2025},
  url={https://arxiv.org/abs/2507.09273}
}

@article{yin-dirac-2026,
  title = {{Nonequilibrium Dynamics of Dirac Quantum Criticality in Imaginary Time}},
  author = {Yu, Yin-Kai and Zeng, Zhi and Shu, Yu-Rong and Li, Zi-Xiang and Yin, Shuai},
  journal = {Phys. Rev. Lett.},
  volume = {136},
  issue = {8},
  pages = {086502},
  numpages = {8},
  year = {2026},
  month = {Feb},
  publisher = {American Physical Society},
  doi = {10.1103/7ltm-f68w},
  url = {https://link.aps.org/doi/10.1103/7ltm-f68w}
}

@article{yin-dirac-natcomm-2025,
  title={{Finite-time scaling beyond the Kibble-Zurek prerequisite in Dirac systems}},
  author={Zeng, Zhi and Yu, Yin-Kai and Li, Zhi-Xuan and Li, Zi-Xiang and Yin, Shuai},
  journal={Nature Communications},
  volume={16},
  number={1},
  pages={6181},
  year={2025},
  publisher={Nature Publishing Group UK London},
  url={https://www.nature.com/articles/s41467-025-61611-6}
}

@article{haldane-spehre-1985,
  title = {{Finite-Size Studies of the Incompressible State of the Fractionally Quantized Hall Effect and its Excitations}},
  author = {Haldane, F. D. M. and Rezayi, E. H.},
  journal = {Phys. Rev. Lett.},
  volume = {54},
  issue = {3},
  pages = {237--240},
  numpages = {0},
  year = {1985},
  month = {Jan},
  publisher = {American Physical Society},
  doi = {10.1103/PhysRevLett.54.237},
  url = {https://link.aps.org/doi/10.1103/PhysRevLett.54.237}
}

@article{jacek2010dynamics,
  title={Dynamics of a quantum phase transition and relaxation to a steady state},
  author={Dziarmaga, Jacek},
  journal={Advances in Physics},
  volume={59},
  number={6},
  pages={1063--1189},
  year={2010},
  publisher={Taylor \& Francis},
  url={https://www.tandfonline.com/doi/full/10.1080/00018732.2010.514702}
}

@article{haldane-pseudo,
  title = {{Fractional Quantization of the Hall Effect: A Hierarchy of Incompressible Quantum Fluid States}},
  author = {Haldane, F. D. M.},
  journal = {Phys. Rev. Lett.},
  volume = {51},
  issue = {7},
  pages = {605--608},
  numpages = {0},
  year = {1983},
  month = {Aug},
  publisher = {American Physical Society},
  doi = {10.1103/PhysRevLett.51.605},
  url = {https://link.aps.org/doi/10.1103/PhysRevLett.51.605}
}

@article{hu-ope-ising-2023,
  title = {{Operator Product Expansion Coefficients of the 3D Ising Criticality via Quantum Fuzzy Spheres}},
  author = {Hu, Liangdong and He, Yin-Chen and Zhu, W.},
  journal = {Phys. Rev. Lett.},
  volume = {131},
  issue = {3},
  pages = {031601},
  numpages = {6},
  year = {2023},
  month = {Jul},
  publisher = {American Physical Society},
  doi = {10.1103/PhysRevLett.131.031601},
  url = {https://link.aps.org/doi/10.1103/PhysRevLett.131.031601}
}

@article{zhou2025-surface-cft,
  title={{Studying the 3d Ising surface CFTs on the fuzzy sphere}},
  author={Zhou, Z and Zou, Y},
  journal={arXiv preprint arXiv:2407.15914},
  year={2025},
  url={https://scipost.org/SciPostPhys.18.1.031}
}

@article{dedushenko2024ising-ising-bcft,
  title={{Ising BCFT from fuzzy hemisphere}},
  author={Dedushenko, Mykola},
  journal={arXiv:2407.15948},
  year={2024},
  url={https://arxiv.org/abs/2407.15948}
}

@article{f-function,
  title = {{Entropic $F$ function of three-dimensional Ising conformal field theory via fuzzy sphere regularization}},
  author = {Hu, Liangdong and Zhu, W. and He, Yin-Chen},
  journal = {Phys. Rev. B},
  volume = {111},
  issue = {15},
  pages = {155151},
  numpages = {9},
  year = {2025},
  month = {Apr},
  publisher = {American Physical Society},
  doi = {10.1103/PhysRevB.111.155151},
  url = {https://link.aps.org/doi/10.1103/PhysRevB.111.155151}
}

@article{spN,
  title={{3D conformal field theories with $Sp(N)$ global symmetry on a fuzzy sphere}},
  author={Zhou, Zheng and He, Yin-Chen},
  journal={Physical Review Letters},
  volume={135},
  number={2},
  pages={026504},
  year={2025},
  publisher={APS},
  url={https://journals.aps.org/prl/abstract/10.1103/xstj-xvcy}
}

@article{zhou2025chern-simens,
  title={{Chern-Simons-matter conformal field theory on fuzzy sphere: Confinement transition of Kalmeyer-Laughlin chiral spin liquid}},
  author={Zhou, Zheng and Wang, Chong and He, Yin-Chen},
  journal={arXiv:2507.19580},
  year={2025},
  url={https://arxiv.org/abs/2507.19580}
}

@article{voinea2025-anyon,
  title={Regularizing 3D conformal field theories via anyons on the fuzzy sphere},
  author={Voinea, Cristian and Fan, Ruihua and Regnault, Nicolas and Papi{\'c}, Zlatko},
  journal={Physical Review X},
  volume={15},
  number={3},
  pages={031007},
  year={2025},
  publisher={APS},
  url={https://journals.aps.org/prx/abstract/10.1103/bf4k-phl9}
}

@article{taige-ON,
  title={{Studying 3D $O(N)$ Surface CFT on the Fuzzy Sphere}},
  author={Feng, Jiechao and Wang, Taige},
  journal={arXiv:2604.21091},
  year={2026},
  url={https://arxiv.org/abs/2604.21091}
}

@article{dey2026conformal-O3,
  title={{Conformal Data for the $O(3)$ Wilson-Fisher Conformal Field Theory from Fuzzy Sphere Realization of the Quantum Rotor Model}},
  author={Dey, Arjun and Herviou, Loic and Mudry, Christopher and L{\"a}uchli, Andreas Martin},
  journal={Physical Review Letters},
  volume={136},
  number={20},
  pages={206504},
  year={2026},
  publisher={APS},
  url={https://journals.aps.org/prl/abstract/10.1103/4yfv-xbcj}
}

@article{guo2025-ON,
  title={The $ {O (N)} $ {Free-Scalar and Wilson-Fisher Conformal Field Theories on the Fuzzy Sphere}},
  author={Guo, Wenhan and Zhou, Zheng and Wei, Tzu-Chieh and He, Yin-Chen},
  journal={arXiv:2512.02234},
  year={2025},
  url={https://arxiv.org/abs/2512.02234}
}

@article{dey2026conformal-O2,
  title={Conformal Data for the $ {O(2)} $ {Wilson-Fisher} {CFT} in $(2+ 1) $-Dimensional {Spacetime} from {Exact} {Diagonalization} and {Matrix} {Product} {States on the Fuzzy Sphere}},
  author={Dey, Arjun and Herviou, Loic and Mudry, Christopher and Rychkov, Slava and L{\"a}uchli, Andreas Martin},
  journal={arXiv:2604.18705},
  year={2026},
  url={https://arxiv.org/abs/2604.18705v2}
}

@article{taylor2026conformal-ising-tricritical,
  title={{Conformal scalar field theory from Ising tricriticality on the fuzzy sphere}},
  author={Taylor, Joseph and Voinea, Cristian and Papi{\'c}, Zlatko and Fan, Ruihua},
  journal={Physical Review Letters},
  volume={136},
  number={5},
  pages={056503},
  year={2026},
  publisher={APS},
  url={https://journals.aps.org/prl/abstract/10.1103/cj3l-cf58}
}

@article{zhou2025free-maj,
  title={{Free Majorana fermion meets gauged Ising conformal field theory on the fuzzy sphere}},
  author={Zhou, Zheng and Gaiotto, Davide and He, Yin-Chen},
  journal={arXiv:2509.08038},
  year={2025},
  url={https://arxiv.org/abs/2509.08038}
}

@article{zohar-2024,
  title={{Impurities with a cusp: general theory and 3d Ising}},
  author={Cuomo, Gabriel and He, Yin-Chen and Komargodski, Zohar},
  journal={Journal of High Energy Physics},
  volume={2024},
  number={11},
  pages={1--47},
  year={2024},
  publisher={Springer},
  url={https://link.springer.com/article/10.1007/JHEP11(2024)061}
}

@article{yin2026finite,
  title={Finite-time Scaling with Arbitrary Driving Rates: Bridging the Kibble-Zurek and De Grandi-Gritsev-Polkovnikov Limits},
  author={Yin, Shuai},
  journal={arXiv:2605.30938},
  year={2026},
  url={https://arxiv.org/abs/2605.30938}
}

@article{shu2026universal,
  title={Universal Short-Imaginary-Time Quantum Critical Dynamics Near Boundaries},
  author={Shu, Yu-Rong and Li, Yuan-Biao and Yin, Shuai},
  journal={arXiv:2607.01076},
  year={2026},
  url={https://arxiv.org/abs/2607.01076}
}

@article{deng2025anomalous,
  title={Anomalous dynamical scaling at topological quantum criticality},
  author={Deng, Menghua and Yang, Sheng and Sun, Chen and Li, Fuxiang and Yu, Xue-Jia},
  journal={arXiv:2512.15537},
  year={2025},
  url={https://arxiv.org/abs/2512.15537}
}

@article{Zhifangxu2005prb,
  title = {Dynamic Monte Carlo renormalization group determination of critical exponents with linearly changing temperature},
  author = {Zhong, Fan and Xu, Zhifang},
  journal = {Phys. Rev. B},
  volume = {71},
  issue = {13},
  pages = {132402},
  numpages = {4},
  year = {2005},
  month = {Apr},
  publisher = {American Physical Society},
  doi = {10.1103/PhysRevB.71.132402},
  url = {https://link.aps.org/doi/10.1103/PhysRevB.71.132402}
}

@article{Gong2010njp,
doi = {10.1088/1367-2630/12/4/043036},
url = {https://dx.doi.org/10.1088/1367-2630/12/4/043036},
year = {2010},
month = {apr},
publisher = {},
volume = {12},
number = {4},
pages = {043036},
author = {Shurong Gong and Fan Zhong and Xianzhi Huang and Shuangli Fan},
title = {Finite-time scaling via linear driving},
journal = {New Journal of Physics}
}

@article{Yin2014open,
  title = {Nonequilibrium quantum criticality in open systems: The dissipation rate as an additional indispensable scaling variable},
  author = {Yin, Shuai and Mai, Peizhi and Zhong, Fan},
  journal = {Phys. Rev. B},
  volume = {89},
  issue = {9},
  pages = {094108},
  numpages = {9},
  year = {2014},
  month = {Mar},
  publisher = {American Physical Society},
  doi = {10.1103/PhysRevB.89.094108},
  url = {https://link.aps.org/doi/10.1103/PhysRevB.89.094108}
}

@article{Zurek2005prl,
  title = {Dynamics of a Quantum Phase Transition},
  author = {Zurek, Wojciech H. and Dorner, Uwe and Zoller, Peter},
  journal = {Phys. Rev. Lett.},
  volume = {95},
  issue = {10},
  pages = {105701},
  numpages = {4},
  year = {2005},
  month = {Sep},
  publisher = {American Physical Society},
  doi = {10.1103/PhysRevLett.95.105701},
  url = {https://link.aps.org/doi/10.1103/PhysRevLett.95.105701}
}

@article{Pol2005prl,
  title = {Universal adiabatic dynamics in the vicinity of a quantum critical point},
  author = {Polkovnikov, Anatoli},
  journal = {Phys. Rev. B},
  volume = {72},
  issue = {16},
  pages = {161201(R)},
  numpages = {4},
  year = {2005},
  month = {Oct},
  publisher = {American Physical Society},
  doi = {10.1103/PhysRevB.72.161201},
  url = {https://link.aps.org/doi/10.1103/PhysRevB.72.161201}
}

@article{Fischer2007prl,
  title = {Vortex Quantum Creation and Winding Number Scaling in a Quenched Spinor Bose Gas},
  author = {Uhlmann, Michael and Sch\"utzhold, Ralf and Fischer, Uwe R.},
  journal = {Phys. Rev. Lett.},
  volume = {99},
  issue = {12},
  pages = {120407},
  numpages = {4},
  year = {2007},
  month = {Sep},
  publisher = {American Physical Society},
  doi = {10.1103/PhysRevLett.99.120407},
  url = {https://link.aps.org/doi/10.1103/PhysRevLett.99.120407}
}

@article{huangyy2014prb,
  title = {Kibble-Zurek mechanism and finite-time scaling},
  author = {Huang, Yingyi and Yin, Shuai and Feng, Baoquan and Zhong, Fan},
  journal = {Phys. Rev. B},
  volume = {90},
  issue = {13},
  pages = {134108},
  numpages = {15},
  year = {2014},
  month = {Oct},
  publisher = {American Physical Society},
  doi = {10.1103/PhysRevB.90.134108},
  url = {https://link.aps.org/doi/10.1103/PhysRevB.90.134108}
}

@article{Liu2014prb,
  title = {Dynamic scaling at classical phase transitions approached through nonequilibrium quenching},
  author = {Liu, Cheng-Wei and Polkovnikov, Anatoli and Sandvik, Anders W.},
  journal = {Phys. Rev. B},
  volume = {89},
  issue = {5},
  pages = {054307},
  numpages = {19},
  year = {2014},
  month = {Feb},
  publisher = {American Physical Society},
  doi = {10.1103/PhysRevB.89.054307},
  url = {https://link.aps.org/doi/10.1103/PhysRevB.89.054307}
}

@article{Sandvik2011prb,
  title = {Universal nonequilibrium quantum dynamics in imaginary time},
  author = {De Grandi, C. and Polkovnikov, A. and Sandvik, A. W.},
  journal = {Phys. Rev. B},
  volume = {84},
  issue = {22},
  pages = {224303},
  numpages = {8},
  year = {2011},
  month = {Dec},
  publisher = {American Physical Society},
  doi = {10.1103/PhysRevB.84.224303},
  url = {https://link.aps.org/doi/10.1103/PhysRevB.84.224303}
}
\appendix
\section*{End Matter}
\section{Slow and fast quench results using ED}
The slow quench limit is more challenging for larger sizes or using TDVP, so that in 
Fig.~\ref{fig:slow} we show  ED results. The expected adiabatic scaling can already be seen \cite{de_grandi_quench_2010,liu_dynamic_2014}.
\begin{figure}[!htbp]
    \centering
    \includegraphics[width=\linewidth]{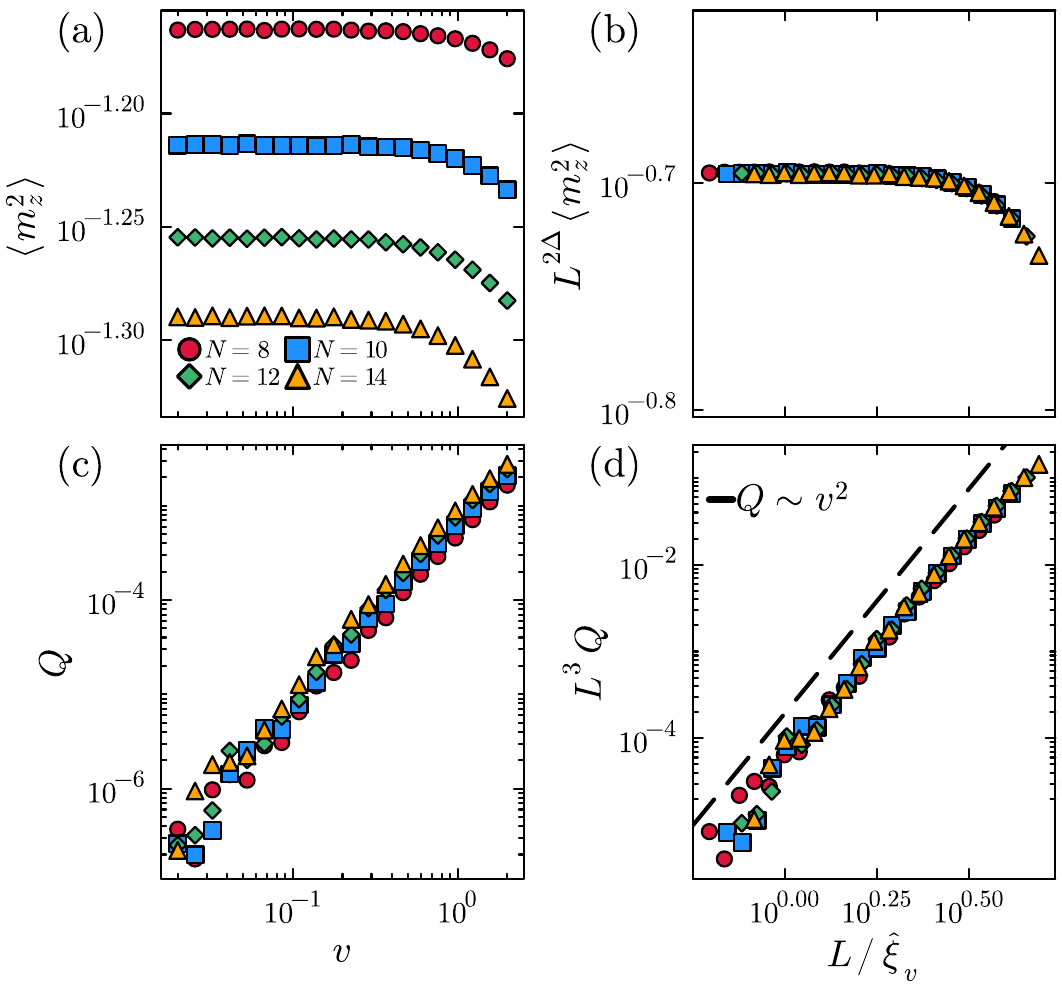}
    \caption{Slow quench regime.}
    \label{fig:slow}
\end{figure}
Fig.~\ref{fig:fast} shows  ED results in the fast quench limit. In particular, Fig.~\ref{fig:fast}(b) shows $\langle m_z^2\rangle\sim L^{-2}$ for sufficiently large quench rate, agreeing with results on conventional lattice \cite{liu_dynamic_2014}. At large $v$, local physics is controlled by a small length scale $\hat{\xi}_v$, making the excitation energy density $Q$ insensitive to system size, shown in Fig.~\ref{fig:fast}(c).
\begin{figure}[!htbp]
    \centering
    \includegraphics[width=\linewidth]{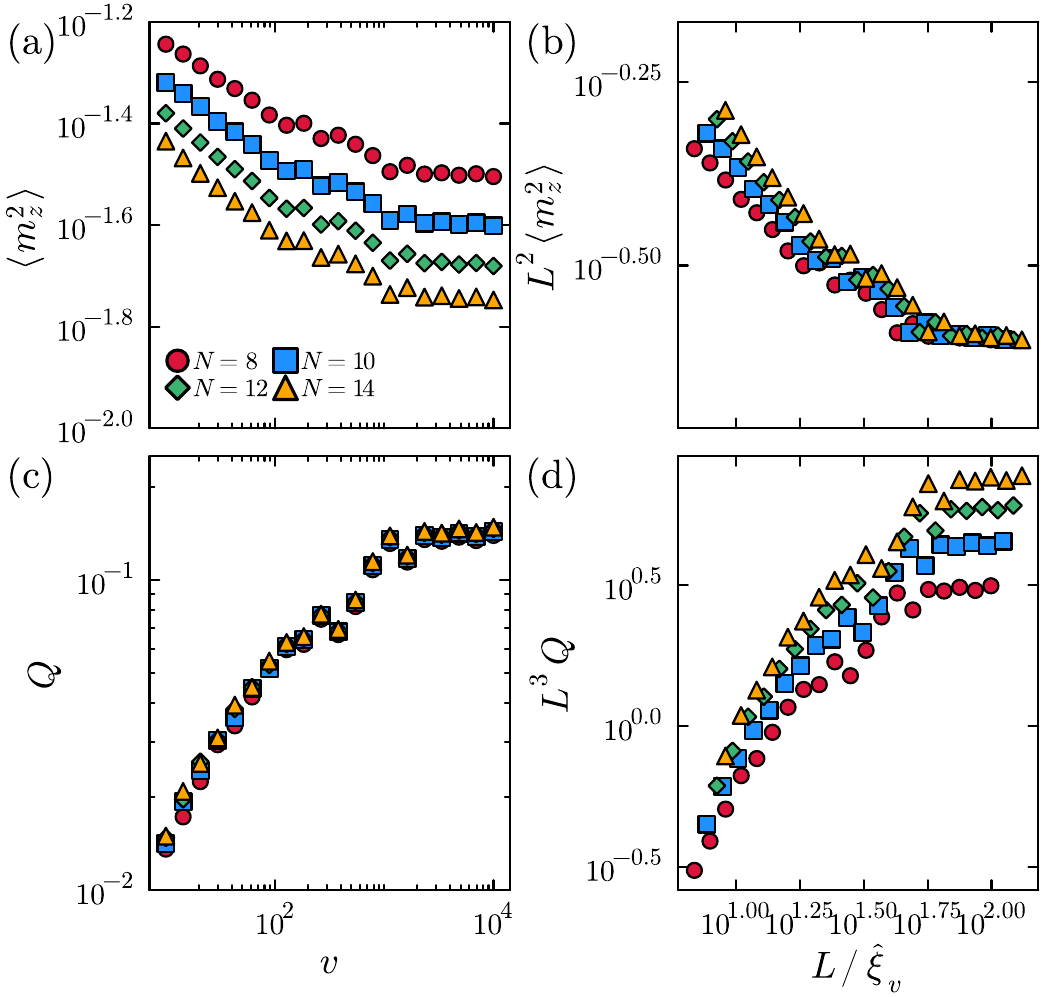}
    \caption{Fast quench regime.}
    \label{fig:fast}
\end{figure}
\section{Level sparsity from symmetry constraint}
\begin{figure}[h]
    \centering
    \includegraphics[width=\linewidth]{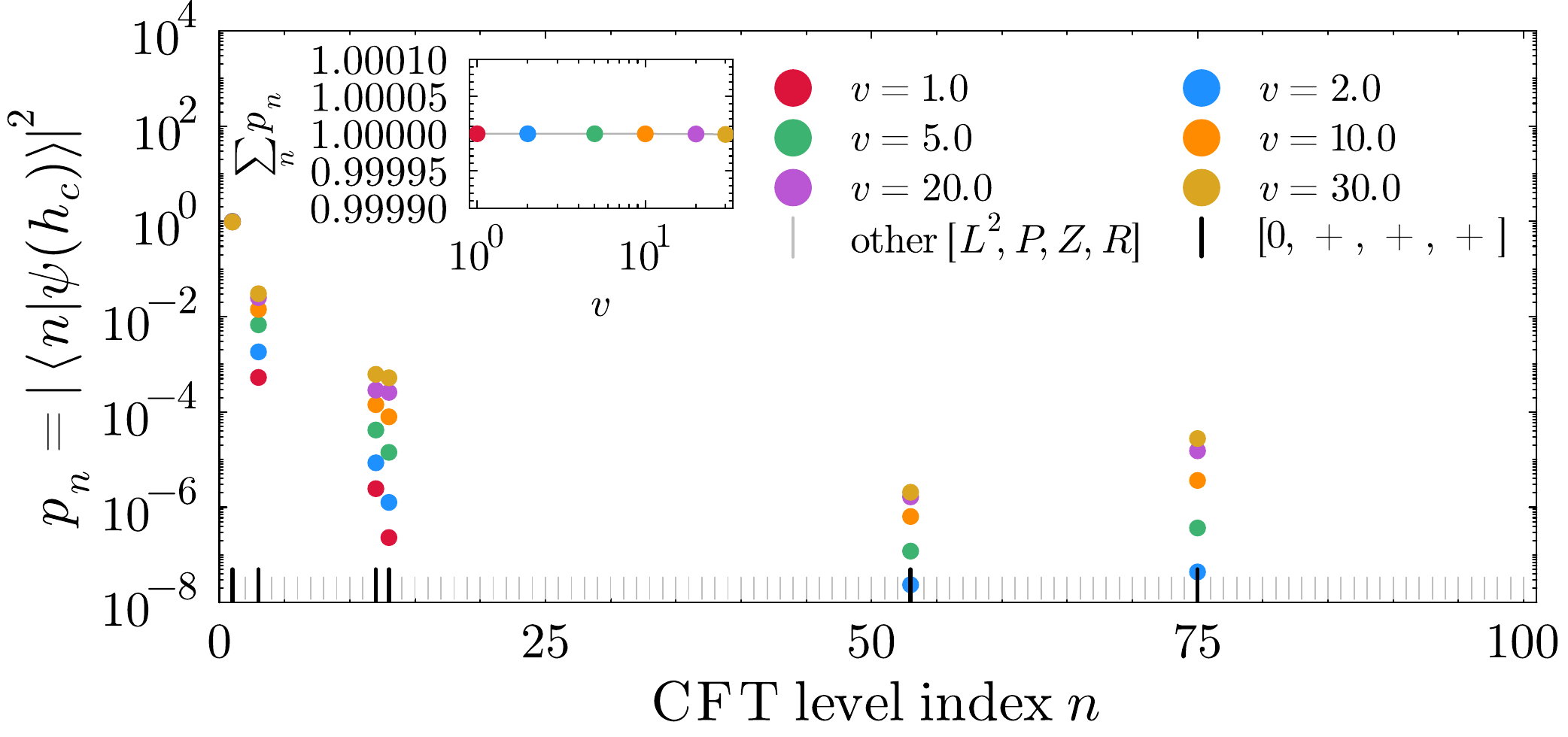}
    \caption{Overlap of the time-evolved state with the 100 lowest CFT levels at the critical point for $N=8$, showing symmetry-enforced level sparsity. Inset: total probability within the 100 levels.}
    \label{fig:excitation}
\end{figure}
In this section, we show the sparsity of the accessible levels using $N=8$, for which the CFT spectrum is readily obtainable. Fig.~\ref{fig:excitation} shows the probability of occupying the $n$-th level for the TDVP evolved state evaluated at the critical point $p_n=|\langle n|\psi(h_c)\rangle|^2$ for a few selected quench rates. The four symmetry numbers $[l,P,Z,R]$ correspond to the angular momentum, the particle-hole (spacetime parity of CFT), the Ising $\mathbb{Z}_2$, and $\pi$-rotation around $y$-axis \cite{zhu_uncovering_2023}. The initial state we used is in the $[0,+,+,+]$ sector, indicated by the thick lines among all the levels. The sparse accessible levels are clearly seen from this figure. For this small system with large finite-size gap $\sim 1/L\sim1/\sqrt{N}$ at the critical point, the time-evolved state occupies predominantly the ground state and has a comparably small probability of occupying higher energy levels. Due to the slow power law decay of the gap with system size, we can conclude that for the system sizes studied in this work, the dynamics is dominated by a few symmetry-allowed levels.
\section{Benchmark and convergence tests}
Fig.~\ref{fig:benchmark} shows the benchmark of the time-evolution algorithm used in this work with ED calculations for a small systems size with $N=8$ electrons. Fig.~\ref{fig:bond-dim} shows a maximum bond dimension $D_\mathrm{max}=1000$ is sufficient to ensure the convergence of all the observables considered. Fig.~\ref{fig:delta-t} shows the ED calculations for $N=8$ with varying time-step $\delta t$ for a quench rate of $v=1$ (the slowest rate used in TDVP simulation in the main text). The results indicate $\lesssim 3\%$ error for $Q$ and $\sim 0.1\%$ error for $\langle m^2_z\rangle$ with a time-step $\delta t=0.05$, which is used in this work.
\begin{figure}[t]
    \centering
    \includegraphics[width=\linewidth]{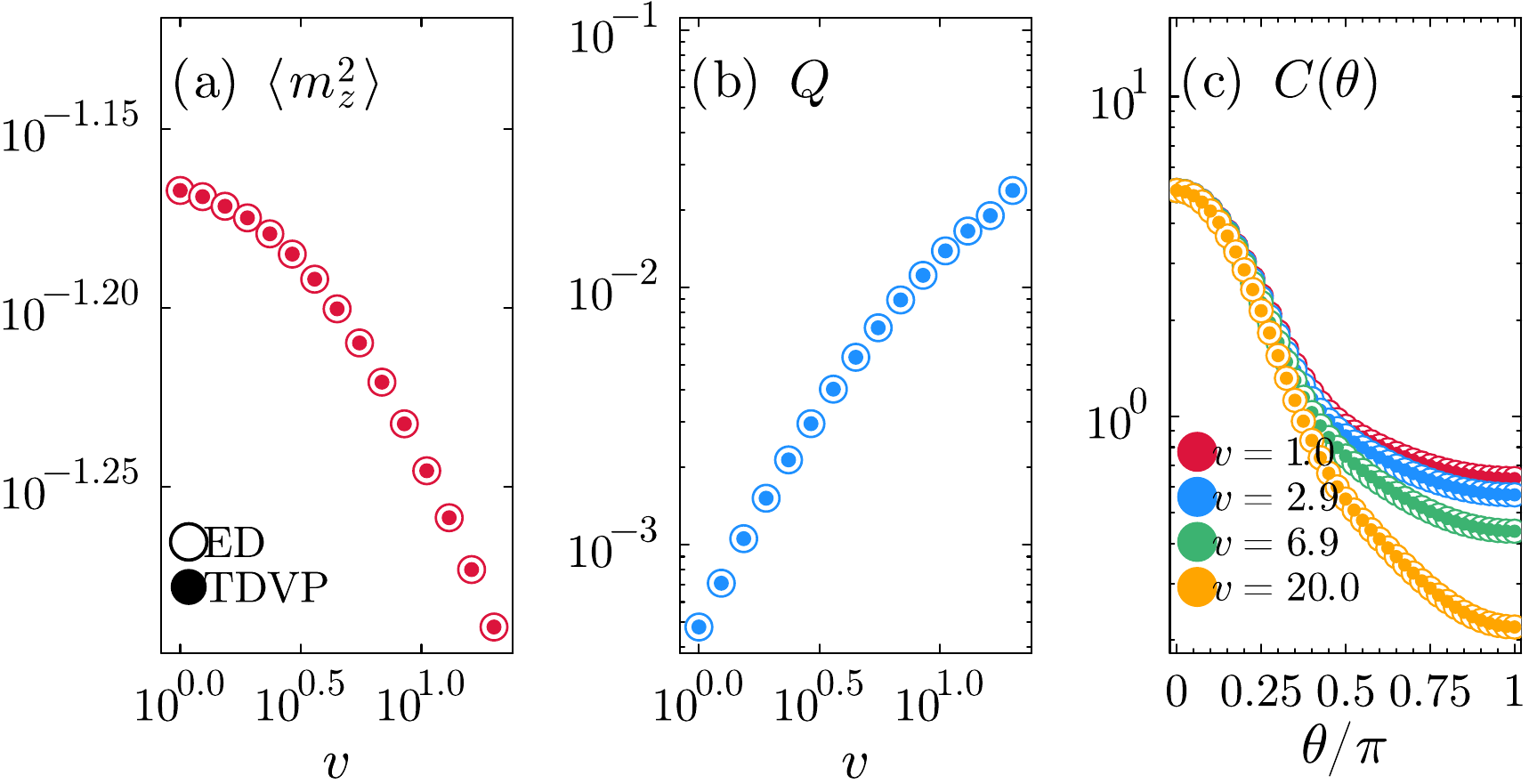}
    \caption{Benchmark of the TDVP results with ED for $N=8$, where the correlation function is shown for four  quench rates.}
    \label{fig:benchmark}
\end{figure}
\begin{figure}[t]
   \centering
   \includegraphics[width=0.75\linewidth]{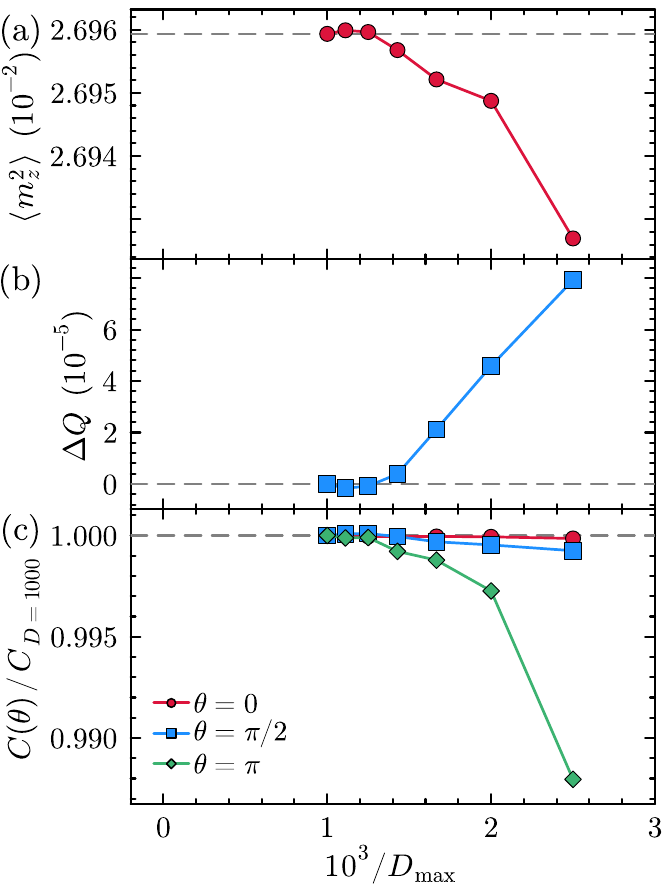}
   \caption{Convergence test of all the three observables using maximum bond dimension from 400 to 1000. $\Delta Q$ in (b) is defined as $Q-Q_{D_\mathrm{max}=1000}$.}
   \label{fig:bond-dim}
\end{figure}
\begin{figure}[t]
    \centering
    \includegraphics[width=0.7\linewidth]{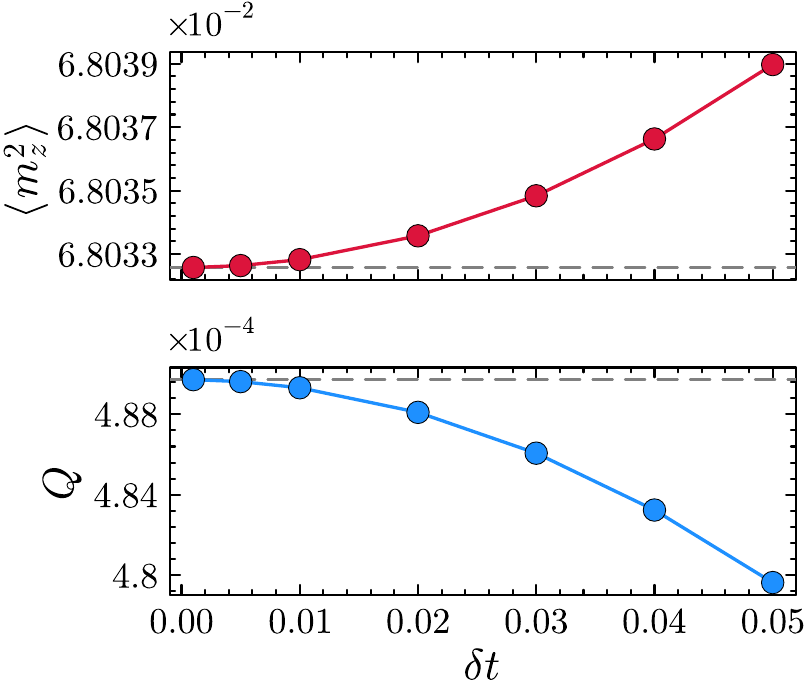}
    \caption{Convergence of $\langle m_z^2\rangle$ and $Q$ when tuning the time-step $\delta t$ for time-evolution. Here $N=8$ and $v=1$.}
    \label{fig:delta-t}
\end{figure}
\clearpage
\onecolumngrid

\end{document}